\documentclass[12pt]{article}%
\usepackage{ amsmath, graphics, rotating}%

\begin{document}
\large
\begin{center} {\bf Tensor Potential Description of Matter and Space, II 
Semi-guage and the Law of Conservation.} \end{center}
\begin{center}
{Boris Hikin}
\small

E-mail:  boris.hikin@verizon.net

Tel: 310-922-4752, or 310-826-0209, USA \end{center}
\vskip 1em
\large
{\it Abstract:}
\vskip 0.5em
\small

Considered a unified field theory approach describing matter and space (metric tensor) by means of a 3-index tensor $P^{\,i}_{jk}$. 
It is shown that if the Lagrangian has partial U(1) gauge (semi-gauge) of the type $P^{\,i}_{jk}->P^{\,i}_{jk}+a \delta{^i_j} \phi{_{,k}}+b \delta{^i_k} 
\phi{_{,j}}+cg_{jk}g^{mi} \phi{_{,m}}$ where a,b,c are constants, then the Euler equations
of motion contain the covariant low of conservation in the form $J^{\,k}_{;k}=0$.

\baselineskip 1pc
\large
\vskip 3em
{\underline {Law of Conservation}}
\vskip 1em

In the first paper \cite{i1}  on this subject we proposed a concept of describing the matter (including the gravitational matter) and the 
space by means of one entity - 3-index tensor (called tensor potential) $P^{\,i}_{jk}$ with no a priori set symmetries. The metric tensor $g_{ij}$
is defined  as a quadratic function of $P^{\,i}_{jk}$.

\begin{eqnarray}
\label{f101}
g_{ij} = b_1{\bar P}^{\, m}_{ni}{\bar P}^{\, n}_{mj}+b_2{\hat P}^{\, m}_{in}{\hat P}^{\, n}_{jm}+
b_3({\bar P}^{\, m}_{in}{\hat P}^{\, n}_{mj}+{\bar P}^{\, m}_{jn}{\hat P}^{\, n}_{mi})+b_4{\bar P}_{\, m}{\bar P}^{\, m}_{ij} \nonumber\\
+\, b_5{\hat P}_m{\hat P}^{\, m}_{ij}+b_6{\bar P}_i{\bar P}_j+b_7{\hat P}_i{\hat P}_j+b_8({\bar P}_i{\hat P}_j+{\bar P}_j{\hat P}_i)
\end{eqnarray}
In the above equation, $\bar P^i_{jk}$ and $\hat P^i_{jk}$ are the symmetric
and antisymmetric (in low indices) parts of the tensor $P^{\, i}_{jk}$, respectively. Vectors $\bar P_k=\bar P^i_{ik}$ and $\hat Pk=\hat P^i_{jk}$
are the vectors obtained by contraction of the tensors $\bar P^i_{jk}$ and $\hat P^i_{jk}$. The coefficients $b_1$ thru $b_8$ are constants.

In \cite{i1} we suggested that $g_{ij}$ depends only on symmetric (low indices) part of $P^i_{jk}$ with these values for 
$b_1$ thru $b_8: b_1=-18,\, b_4=12,\, b_6=1,\, and\, b_i =0$ if i $\neq$ 2,4,6. This choise on parameters $b_1$, ..., $b_8$ 
is only one possibility and it is not relevant to the subject of this paper.

The tensor of matter $M^i_{jkl}$ would be defined as a combination (sum) of covariant derivatives (using $g_{ij}$) of $P^i_{jk}$
($P^i_{jk;l}$) and the square of $P^i_{jk}$. Such structure of the tensor of matter $M^i_{jkl}$ is dictated by the scaling factors of  $P^{\, i}_{jk}$ 
and $M^i_{jkl}$: $P^{\, i}_{jk}$ is $cm^{-1}$ and $M^{\, i}_{jkl}$ is $cm^{-2}$.
Examples of such a tensor matter are shown below:

\begin{eqnarray}
\label{f102}
&&a)M^i_{jkl}= P^{\, i}_{jk;l} \nonumber\\
&&b)M^i_{jkl}= 3P^{\, i}_{jk;l}+P^{\, i}_{mk}P^{\, m}_{jl} \\
&&c)M^i_{jkl}= P^{\, i}_{jl;k}-P^{\, i}_{jk;l}+P^{\, i}_{mk}P^{\, m}_{jl}-P^{\, i}_{ml}P^{\, m}_{jk} \nonumber
\end{eqnarray}

The equation for the distribution of matter is obtained from the variational principle. The Lagrangian depends
on tensors $M^{\, i}_{jkl}$ and $g_{ij}$, which both are functions of only the tensor potential $P^{\, i}_{jk}$.
\begin{equation}
\label{f103}
A=\int\nolimits L(M^{\, i}_{jkl}(P^{\, i}_{jk}, P^{\, i}_{jk;l}), \, g_{ij}(P^{\, i}_{jk} ) ){\,} {\sqrt {g(P^{\, i}_{jk} )}} \,\, d^4x
\end{equation}
where $g(P^{\, i}_{jk} )=-det(g_{ij})$.

By analogy with other flat space field theories, we will assume that the Lagrangian is quadratic with respect to the matter tensor M.
From a physical point of view it is more convenient to consider $P^{\,i}_{jk}$ and $g_{ij}$ as independent variables. 
By introducing Lagrange coefficients
(or Lagrange multipliers) $T^{\, ij}$, we can rewrite the action integral A in this form:

\begin{equation}
\label{f104}
A=\int\nolimits \sqrt{g}{\,} d^4x{\,} \{L(P^{\, i}_{jk;l}, P^{\, i}_{jk}, g_{ij})+T^{\, mn}[g_{mn}-X_{mn}(P^{\, i}_{jk})]\}
\end{equation}

Here, $X_{mn}(P^{\, i}_{jk})$ is a quadratic function given by eq.(\ref{f101}).
The variational principle with respect to $P^i_{jk}$, $g_{ij}$ and $T^{ij}$ yields the following three sets of Euler equations
(for details see \cite{i1}):
\begin{equation}
\label{f105}
\frac {{\delta}A} {{\delta}P^{\, i}_{jk}}=0{\quad}or {\quad}
{\partial}L/{{\partial}P^{\, i}_{jk}}-({\partial}L/{{\partial}P^{\, i}_{jk;l})}_{;{\,}l}-T^{\, mn}[{\partial}X_{mn}/{\partial}P^{\, i}_{jk}]=0
\end{equation}

\begin{equation}
\label{f106}
\frac {{\delta}A} {{\delta}g_{ij}}=0{\quad} or \quad \frac {{\delta}L} {{\delta}g_{ij}}+T^{ij}=0
\end{equation}

\begin{equation}
\label{f107}
\frac { {\delta}A}{{\delta}T^{ij}}=0 {\quad} or {\quad} g_{ij}-X_{ij}(P^i_{jk})=0
\end{equation}
Obviously, the set of Lagrange multipliers $T^{ij}$, defined by eq. (\ref{f106}), is the energy-momentum tensor of the matter.

We now show that if the first part of Lagrangian (function L) has the U(1) gauge of the type
\begin{equation}
\label{f108}
P^i_{jk}->P^i_{jk}+a \delta ^i_j \phi _{,k}+b \delta ^i_k \phi _{,j}+cg_{jk}g^{mi} \phi _{,m}
\end {equation}
where a,b,c are constants, 
then the Euler equations of motion (\ref{f105}) contain the covariant low of conservation in the form ${J^k}_{;k}=0$. 

Let us first consider the action without the "constraint" part
\begin{equation}
\label{f109}
A=\int\nolimits \sqrt{g}{\,} d^4x{\,} \, L(P^{\, i}_{jk;l}, P^{\, i}_{jk}, g_{ij})
\end{equation}

Here $P^i_{jk}$ are the independent variables and $g_{ij}$ are some fixed functions not depending on $P^i_{jk}$
The variation of this integral with respect to the tensor $P^i_{jk}$ can be written in this form:
\begin{equation}
\label{f110}
{{\delta}A}=\int\nolimits \sqrt{g}{\,} d^4x{\,} Q_i^{jk} \, \delta P^i_{jk}
\end{equation}
where
\begin{equation}
\label{f111}
Q_i^{jk}={\partial}L/{{\partial}P^{\, i}_{jk}}-({\partial}L/{{\partial}P^{\, i}_{jk;l})}_{;{\,}l}
\end{equation}

Let us now assume that the Lagrangian L has a guage. That is to say, the Lagrangian does not change when 
we replace $P^i_{jk}$ with $P^i_{jk}+a \delta ^i_j \phi _{,k}+b \delta ^i_k \phi _{,j}+cg_{jk}g^{mi} \phi _{,m}$
,where a,b,c are constants. If we assume that $\delta P^i_{jk}$ is due to change in fuction $\phi$ only, we can now rewrite the variation
of action integral in this form:

\begin{eqnarray}
\label{f112}
&&{\delta A}=\int\nolimits \sqrt{g}{\,} d^4x{\,}Q_i^{jk}\, \delta P^i_{jk}= \nonumber\\
&&\int\nolimits \sqrt{g}{\,} d^4x{\,}Q_i^{jk}\, \delta (a \delta ^i_j \phi _{,k}+b \delta ^i_k \phi _{,j}+cg_{jk}g^{mi} \phi _{,m})=\nonumber\\
&&\int\nolimits \sqrt{g}{\,} d^4x{\,}(-a{Q_i^{ik}}_{;k}-{bQ_i^{ki}}_{;k}-c({Q_k^{ij}g_{ij})}^{;k})\delta \phi
\end{eqnarray}

Since L has the gauge and thus it does not depend on $\phi$, $\delta A$ must equal zero for any $\phi$. From this it follows that 
$(-a{Q_i^{ik}}_{;k}-{bQ_i^{ki}}_{;k}-c({Q_k^{ij}g_{ij})}^{;k})=0$ for any $P^i_{jk}$ or, in other words, it is an identity. 
This in fact is a particular case of the second Noether theorem. The additional illustraton can be found in Appendix A, 
where we give an example of Lagrangian, full derivation of Euler equations and derivation from them the Noether identities.

We now consider a situation were $g_{ij}$ is not a fixed function, but a function of $P^i_{jk}$.
By introducing Lagrange multipliers $T_{ij}$ (Lagrange coefficient), we can consider $g_{ij}$ to be an independent variable
with the Lagrangian represented by (\ref{f104}).

The Euler equation with respect to variation of $P^i_{jk}$ will take this form. 
\begin{eqnarray}
\label{f113}
&&Q_i^{jk}+J_i^{jk}=0 \quad where \nonumber\\
&&Q_i^{jk}={\partial}L/{{\partial}P^{\, i}_{jk}}-({\partial}L/{{\partial}P^{\, i}_{jk;l})}_{;{\,}l}\quad and \nonumber\\
&&J_i^{jk}=T^{\, mn}[{\partial}Q_{mn}/{\partial}P^{\, i}_{jk}]
\end{eqnarray}

If we now construct the invariant the same way as it appears in (\ref{f112}) we will get:
\begin{equation}
\label{f114}
[(-a{Q_i^{ik}}_{;k}-b{Q_i^{ki}}_{;k}-cQ_k^{ij;k}g_{ij})]+[(-a{J_i^{ik}}_{;k}-b{J_i^{ki}}_{;k}-cJ_k^{ij;k}g_{ij})]=0
\end{equation}
The expression in the first square bracket, as it has been shown above, vanishes leaving the remaining equation in the form:
\begin{equation}
\label{f115}
J^k_{;k}=0 \quad where \quad J^k=(aJ_i^{ik}+bJ_i^{ki}+cJ_m^{ij}g_{ij}g^{mk})
\end{equation}
This is the covariant law of conservation. Thus we have shown that if Lagrangian has the guage and the constraints on 
$g_{ij}$ do not (thus semi-gauge), the Euler equations contain the Law of Conservation. The requirement that tensor $M^i_{jkl}$, which makes
up the Lagrangian has this gauge is not necessary as long as Lagrangian by itself has such a gauge.

This situation is very similar to the law of conservation of electrical 4-current in the classical Maxwell electromagnetic theory.
The Lagrangian in this case can be written as:
\begin{equation}
\label{f116}
I=\int\nolimits \sqrt{g}{\,} d^4x{\,} [F^{ij}F_{ij}+A_iJ^i] \quad ,where F_{ij}=A_{i;j}-A_{j;i}
\end{equation}
This Lagrangian has semi-gauge $A_i=A_i+f_{,i}$. This transformation does not change the electromagnetic tensor $F_{ij}$
and thus the first term of the Lagrangian. However, $A_i=A_i+f_{,i}$ is not a gauge for the second term containing the 4-current $J^i$.
As the result, the Maxwell equations contain the law of conservation of 4-current.

It can be seen from eq.(\ref{f113}) that the 4-current $J_k$  in our theory is proportional to $T_{ij}$ or energy-momentum tensor. 
Using (\ref{f101}) $J_k$ can be written directly in terms of $T_{ij}$ and tensor potential $P^i_{jk}$. It will contain only 7 terms as shown below:
\begin{eqnarray}
\label{f117}
&&J_k=\bar b_1T^{mn}\bar P_{mkn}+\bar b_2T^{mn}\hat P_{mkn}+\bar b_3T^{mn}\bar P_{kmn}\nonumber\\
&&+\bar b_4T^m_k\bar P_m + \bar b_5T^m_k\hat P_m +\bar b_6T\bar Pk+\bar b_7T\hat Pk
\end{eqnarray}
Here $\bar b_1$ ... $\bar b_8$ are constants that depend on constants $b_1$ ... $b_8$ of eq.(\ref{f113}) and gauge constant a,b,c.

In \cite{i1}, we showed that contraction of Euler equations obtained by variation of 
$g_{ij}$ ($g^{ij} \delta L/ \delta g_{ij})$
leads to the equation:
\begin{equation}
\label{f118}
\bar J^k_{;k}+T=0 \quad where \quad T=T^i_i
\end{equation}
This is a direct consequence of the fact that Lagrangian is a quadratic function of the tensor of matter $M^i_{jkl}$.
In order to yield the law of conservation, we had to postulate that for all physically meaningful solutions the invariant T (the 
trace of $T_{ij}$) for entire system (including gravitational field) is zero. 
If the Lagrangian has a semi-gauge, this requirement is no longer needed.
However, if in some case T=0, we would have one more 4-current that is conserved.

\baselineskip 1pc
\large
\vskip 3em
{\underline {Lagrangian}}
\vskip 1em

We now discuss the question of the freedom that we have in the choice of Lagrangian with such semi-gauge.
It is well known that if the tensor ${M^{\,i}}_{jkm}$ is defined by expression (\ref{f102}c), the Lagrangian (\ref{f103})
satisfies the gauge of $P^i_{jk}->P^i_{jk}+\delta ^i_j \phi_{,k}$. 
We will show now that the requirement for Lagrangian having the more general gauge (\ref{f108}) 
$P^i_{jk}-> P^i_{jk}+a \delta ^i_j \phi _{,k}+b \delta ^i_k \phi _{,j}+cg_{jk}g^{mi} \phi _{,m}$
is not that difficult to satisfy.

Let us consider Lagrangians that depend only on $P^i_{jk;m}$. In general there are total of 24 (4!) terms in which 4 indices
of one tensor P are contracted with 4 indices of the other tensor P. Here are a couple of examples of such terms: 
$P_{ijkm}P^{ijkm}$ or $P_{ijkm}P^{mijk}$ - where $P_{ijkm}=P_{ijk;m}=g_{is}P^s_{jk;l}$ 
In the expression above, as in all similar expressions below, the sign of  semicolon (;) is dropped.
Only 17 out of 24 quadratic scalars are independent. Below we give a complete list of these invariants. 
The first group represents the terms that are symmetrical with respect to renaming indeces. 
Meaning, if we rename the indices of the second P to be "ijkm" , the term transfers into itself. For example:
\begin{equation}
P^{ijkm}P_{mkji}=_{(m->i, k->j, j->k, i->m} =P^{mkji}P_{ijkm}=P^{ijkm}P_{mkji}
\end{equation}

There are 10 such terms: 
\begin{eqnarray}
\label{f119}
P_{ijkm}P^{ijkm}, \, P_{ijkm}P^{ijmk},\, P_{ijkm}P^{imkj},\, P_{ijkm}P^{ikjm},\, P_{ijkm}P^{jikm},\, \nonumber\\
P_{ijkm}P^{jimk},\, P_{ijkm}P^{kmij},\, P_{ijkm}P^{kjim},\, P_{ijkm}P^{mjki},\, P_{ijkm}P^{mkji}\, 
\end{eqnarray}

The second group contains 14 terms in which only 7 are independent with respect to renaming indices. 
\begin{eqnarray}
\label{f120}
&&P_{ijkm}P^{ikmj}=_{(i->i\,,\, k->j\,,\, m->k\,,\, j->m)}=P_{imjk}P^{ijkm}=P_{ijkm}P^{imjk} \nonumber\\
&&P_{ijkm}P^{jkmi}=_{(j->i\,,\, k->j\,,\, m->k\,,\, i->m)}=P_{mijk}P^{ijkm}=P_{ijkm}P^{mijk} \nonumber\\
&&P_{ijkm}P^{jmik}=_{(j->i\,,\, m->j\,,\, i->k\,,\, k->m)}=P_{kimj}P^{ijkm}=P_{ijkm}P^{kimj} \nonumber\\
&&P_{ijkm}P^{jmki}=_{(j->i\,,\, m->j\,,\, k->k\,,\, i->m)}=P_{mikj}P^{ijkm}=P_{ijkm}P^{mikj} \nonumber\\
&&P_{ijkm}P^{jkim}=_{(j->i\,,\, k->j\,,\, i->k\,,\, m->m)}=P_{kijm}P^{ijkm}=P_{ijkm}P^{kijm} \nonumber\\
&&P_{ijkm}P^{kjim}=_{(k->i\,,\, j->j\,,\, i->k\,,\, m->m)}=P_{kjim}P^{ijkm}=P_{ijkm}P^{kjim} \nonumber\\
&&P_{ijkm}P^{kmji}=_{(k->i\,,\, m->j\,,\, j->k\,,\, i->m)}=P_{mkij}P^{ijkm}=P_{ijkm}P^{mkij} \nonumber\\
\end{eqnarray}

In addition to the terms discussed above, there are terms where the contraction of indices occurs once inside $P_{ijkm}$.
An example of one such term is ${{P^i}_i}_{km}{P_j}^{jkm}$

In general, the contraction of tensor $P_{ijkm}$ will produce six 2-index tensors: $P^{(1)}_{km}={P^i}_{ikm}$, 
$P^{(2)}_{jm}={P^i}_{jim}$, $P^{(3)}_{jk}={P^i}_{jki}$, $P^{(4)}_{im}=P_{ijkm}g^{jk}$, $P^{(5)}_{ik}=P_{ijkm}g^{jm}$,
$P^{(6)}_{ij}=P_{ijkm}g^{km}$.
If the symmetry of these tensors is not defined (meaning, the tensor has both symmetric and antisymmetric parts),
there would be 21 Lagrangian terms that can be constructed. These Lagrangian terms could be written in this form:

\begin{equation}
\label{f121}
P^{(\alpha)}_{km}P^{(\beta)km}\,,\, P^{(\alpha)}_{km}P^{(\beta)mk}
\end{equation}
where $\alpha$ and $\beta$ takes values from 1 to 6 and $\alpha>=\beta$

The six 2-index tensors $P^{(\alpha)}_{km}$ ($\alpha=1,...,6$) allow to form 3 scalars: $I_1={{P^{\,i}}_{ik}}^k$,
$I_2={{P^{\,i}}_{ki}}^k$, $I_3={{{P^{\,i}}_k}^k}_i$.
Using these 3 scalars, 6 more Lagrangian terms can be written as a product of the 2 scalers: 

\begin{equation}
\label{f122}
I_1I_1\,,\,I_1I_2\,,\,I_1I_3\,,\,I_2I_2\,,\,I_2I_3\,,\,I_3I_3
\end{equation}

Combining these three types of Lagrangian terms together (eq.(\ref{f119}) thru eq.(\ref{f122})),
 we will have the total of 44 terms or 44 unknown constants
(out of which, of course, only 43 are independent).

In order to evaluate the ability of Lagrangian to have the gauge of type (\ref{f101}), we have to replace
 ${P^{\,i}}_{jk}$ with ${P^{\,i}}_{jk}+a \delta ^i_j \phi _{,k}+b \delta ^i_k \phi _{,j}+cg_{jk}g^{mi} \phi _{,m}$
and then evaluate the requirement on these 44 constants to assure that the final Lagrangian does not contain any terms
that include function $\phi$.
It is not diifcult to see that due to the symmetry of $\phi_{\,;i;j}$ ($\phi_{\,;i;j}=\phi_{\,;j;i}$)  there are only 11 such terms.

\begin{eqnarray}
\label{f123}
&&P^{(\alpha)km}\phi_{\,;k;m} \quad \alpha=1\,,...\,,6\\
&&I_1\phi_{\,;k;m}g^{km}\,,\,I_2\phi_{\,;k;m}g^{km}\,,\,I_3\phi_{\,;k;m}g^{km}\\
&&\phi_{\,;i;j}\phi_{\,;k;m}g^{ik}g^{jm}\,\, and\,\, \phi_{\,;i;j}\phi_{\,;k;m}g^{ij}g^{km}
\end{eqnarray}

Thus for any pre-set values of constants a,b,c in the gauge (\ref{f108}) we will have 11 linear equations 
with 44 unknown constants associated with all possible Lagrangian terms.

The reason why the number of equations is much less then the number of possible Lagrangianian terms is due
to the fact that the gauge contains the metric $g_{ij}$.
The presence of the metric tensor reduces the rank of ${P^{\,i}}_{jk}$ tensor and thus the number of possible
terms that contain gauge function $\phi$.

The same can be observed for the other types of Lagrangianian terms - those that consist of ${P^{\,i}}_{jk;m}$
and the square of ${P^{\,i}}_{jk}$ (an example of such a term could be ${P^{\,i}}_{jk;m}{P^{\,n}}_{np}{P^{\,j}}_{iq}g^{kp}g^{mq}$)
or the ones that consist of four-product of ${P^{\,i}}_{jk}$  (example - $P_{ijk}P^{imn}P^{jsk}{P^s}_{ms}g_{ns}$).

Thus we conclude that the requirement of Lagrangian to have a gauge of type (\ref{f108}) is quite possible to satisfy.

\baselineskip 1pc
\large
\vskip 3em
{\underline {Appendix A}}
\vskip 1em

Here we will illustrate how semi-gauge leads to the law of conservation using a particular example.
Let's assume that $M^i_{jkl}$ is defined in terms of $P^i_{jk}$ by formula (\ref{f102}c):

\begin{equation}
\label {f124}
M^i_{jkl}= P^{\, i}_{jl;k}-P^{\, i}_{jk;l}+P^{\, i}_{mk}P^{\, m}_{jl}-P^{\, i}_{ml}P^{\, m}_{jk}
\end{equation}

It is not difficult to show that $M^i_{jkl}$  has a guage with respect to the following transformation of 
$P^i_{jk}-> P^i_{jk}+\delta ^i_j \phi _{,k}$

Since $M^i_{jkl}$  has a gauge, so will a Lagrangian that depends on $M^i_{jkl}$  only. 
For example if we take $L=M^i_{jkl}*M^j_{imn}g^{km}g^{ln}$, the expression for Euler equations can be written as this:

\begin{eqnarray}
\label{f125}
&&Q_i^{jk}=0\quad,\, where\nonumber\\
&&{Q_i}^{jk}=-2[{{M_i}^{jkl}}_{;l}+{M_i}^{nkl}{P^j}_{nl}-{M_n}^{jkl}{P^{\,n}}_{il}]
\end{eqnarray}

Contracting first 2 indices we will have
\begin{equation}
\label {f126}
{Q_i}^{ik}=-2[{{M_i}^{ikl}}_{;l}+{M_i}^{nkl}{P^i}_{nl}-{M_n}^{ikl}{P^{\,n}}_{il}]={N^{kl}}_{\,;l}
\end{equation}
where $N^{kl}={M_i}^{ikl}$ is antisymmetric tensor due to the antisymmetry of ${M_i}^{jkl}$ in indices k,l.
If we take the covariant derivative of ${Q_i}^{ik}$ and contract it by index k (${{Q_i}^{ik}}_{;k}$), we will get:

\begin{eqnarray}
\label{f127}
&&{{Q_i}^{ik}}_{;k}={N^{km}}_{;k;m} =1/2({N^{km}}_{;m;k}-{N^{km}}_{;k;m})=\nonumber\\
&&1/2({R_{kms}}^kN^{sm}+{R_{kms}}^mN^{ms})=R_{ms}N^{sm}\equiv 0
\end{eqnarray}

The last expression is an identity since tensor Ricci  ($R_{ij}$) is symmetric and tensor $N_{ij}$ is antisymmeric.

Again this fact of identity is direct consequence of second Noether theorem and is true not only for this example, 
but for any case when Lagrangian has a gauge.

\end{document}